\documentclass[aps,amsmath,amsfonts,11pt]{revtex4}
\usepackage{graphicx}
\usepackage{epstopdf}
\usepackage{epsfig,graphicx}
\usepackage[english]{babel}
\usepackage{amsfonts}
\usepackage{amsmath}
\usepackage{latexsym}
\usepackage{graphics,bm}
\usepackage{dcolumn}
\usepackage{bm}
\usepackage{rotating}

\begin{document}

\title{Negative effective mass, optical multistability and Fano line-shape control via mode tunneling in double cavity optomechanical system}

\author{ Vaibhav N Prakash$^{1}$ and Aranya B Bhattacherjee$^{2}$ }

\address{$^{1}$School of Physical Sciences, Jawaharlal Nehru University, New Delhi-110067, India} 
\address{$^{2}$Department of Physics, Birla Institute of Technology and Science, Pilani,
Hyderabad Campus,  Hyderabad - 500078, India}

\begin{abstract}
We theoretically investigate the optical and mechanical properties of a double cavity optomechanical system with one stationary and two harmonically bound mirrors. We show that it is possible for the mechanical 
mirrors in this system to posses negative effective mass. Working within the strong coupling and the resolved sideband regime, we show that the displacement of the middle resonator is multistable under certain constrains. We also point to the existence of optomechanically induced absorption (OMIA) and Fano resonance. Owing to the negative effective mass, our scheme can be exploited in the study of quantum optomechanical metamaterials.
\end{abstract}

\maketitle

\section{Introduction}
Questions regarding the foundations of quantum mechanics, the storage and transfer of quantum information and sensitive detector technologies have made quantum optomechanics
a field of considerable interest.  The field has undergone rapid development over the last decade with the development of new methods for fabricating small bulk
mechanical resonators of various forms; nanoscale beams coupled to microwave cavities \citep{1,2,3,4,5,6}, photonic-phononic crystals \citep{7,8,9,10,11}, toroidal optical micro-resonators \citep{12,13,14,15,15a}, doubly
clamped beams with integrated mirrors and drumhead capacitors in superconducting microwave resonators \citep{16,17,18,19}.\\
In recent years another interesting field of study related to acoustic and optical metamaterials has emerged and developed quite rapidly. These metamaterials can be engineered to posses 
properties such as negative refractive index or enhanced absorption below and above a certain cutoff frequency\citep{20,10}. They  are hence dynamic in nature, as opposed to the known static 
properties. Due to their peculiar properties, they can be used as cloaking devices or high-pass and low-pass filters. At the heart of these metamaterials is the ability to manipulate 
acoustic or electromagnetic waves due to properties such as negative effective inertia(negative effective density)\citep{21,22,23} and/or negative bulk modulus/compressibility (in case of acoustic
waves). One of the key features of acoustic and optical metamaterials is that they posses a dynamic negative effective mass.\\
The coupling of optical modes to those of the mechanical oscillator, say for instance in a simple Fabry Perot cavity with a movable end mirror, introduces non-linearity in the system. 
At negative cavity detuning, $\Delta<0$ ($\Delta=\omega_{laser}-\omega_{cav}$), this coupling can be used to slow down the motion of the oscillator. This 'optical' spring effect can be used to alter 
the simple harmonic potential of the oscillator to have bistable or multistable behavior\citep{24,25,26}. Stability is an important aspect when it comes to all optical switches in fibre optic 
communications. Another interesting phenomena in opto-mechanics  is ``OptoMechanically Induced Transparency(OMIT)''\citep{27,28,29,30,31}, caused due to destructive interference between intracavity 
field and the weak laser probe field. It is similar to Electromagnetically Induced Transparency(EIT) often seen in quantum optics.\\

In this article we look at the steady state behavior and fluctuation dynamics inside a Fabry Perot cavity with two movable mirrors coupled via cavity field. We calculate multistability 
and also show OMIA behavior at zero tunneling rate ($g=0$) of photons from one cavity into the other. We show that asymmetric Fano lineshape exist where the probe frequency matches the intracavity frequency at nonzero tunneling rates. We attempt to demonstrate that mechanical resonators coupled to cavity fields can posses a negative effective mass. This might have far reaching consequences in the field of quantum optomechanics.
It is important to note that all experimental parameters used in this article are technologically accessible.\\
The paper is organized as follows. In Section 2, we describe the Hamiltonian of the system followed by the quantum Langevin equations for various operators. In Section 3 we indicate 
the conditions under which the mechanical resonators attain a negative and zero effective masses. In section 4 and 5 we analyze multistability of the mean displacement of the middle 
mechanical oscillator coupled to the  intracavity photons and calculate the normalized forward transmission and backward reflection of the probe field power. Finally all necessary calculations are given in Appendix-A and B.

\section{Theoretical model}
Our model consists of a Fabry Perot cavity with three mirrors out of which two are movable and one is fixed, as shown in Fig.1. The fixed mirror is semi-transparent allowing the laser light to penetrate inside the cavity while the end mirror is completely opaque to the cavity field. The middle movable mirror  is semi-transparent, the transparency of which can be
adjusted, giving rise to the tunable tunnelling coefficient $g$. Thus cavity A is driven by an intense pump/control laser of frequency $\omega_{c}$ and has an average photon number $\overline{n}_a=<a^{\dagger}a>$ , where $a$ ($a^{\dagger}$) is the annihilation (creation) operator of optical mode confined in cavity A.

\begin{figure}[h]
    \centering
    \includegraphics[scale=0.25]{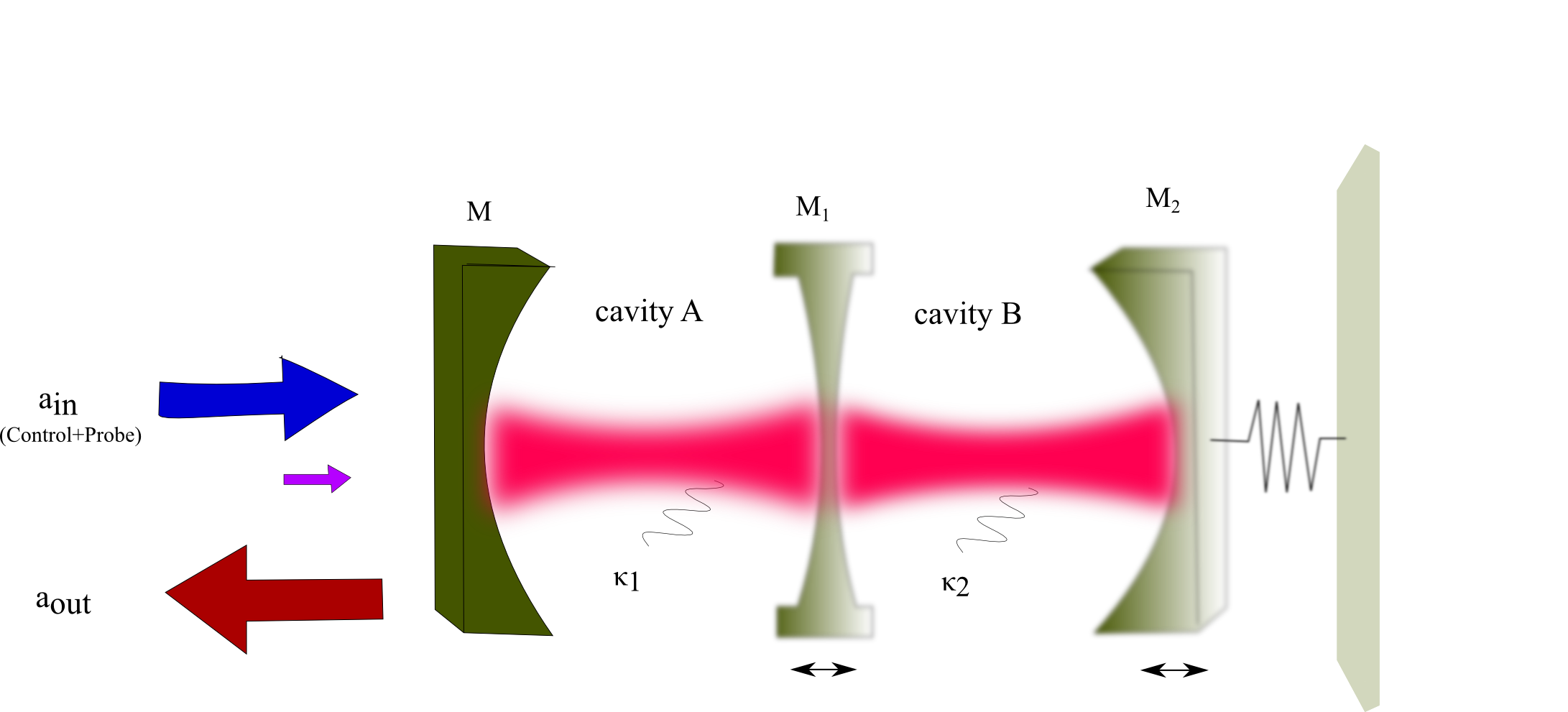}
    \caption{Schematic diagram of double cavity system}
    \end{figure}

The adjustable transparency of the middle mirror allows for the tunnelling of photons with a tunneling rate 'g' into cavity B which has photons with average photon number $\overline{n}_b=<b^{\dagger}b>$  where $b$ ($b^{\dagger}$) is the annihilation (creation) operator of optical mode confined in cavity B. This leads to the coupling of the two mirrors $M_1$ and $M_2$  via radiation pressure forces from the cavity fields. In the frame rotating with frequency $\omega_c$ the Hamiltonian of the hybrid system can be written as,

\begin{equation}
\begin{aligned}
\hat{H}=\sum_{j=1}^{2}(\frac{p_j^2}{2m_{j}}+\frac{1}{2}m_{j}\Omega_j^2\hat{x}_{j}^2)-\hbar\Delta_{1}\hat{a}^\dagger\hat{a}-\hbar\Delta_{2}\hat{b}^\dagger\hat{b}-\hbar G_{1}\hat{x}_{1}\hat{a}^\dagger\hat{a}-\hbar G_{2}(\hat{x}_{2}-\hat{x}_{1})\hat{b}^\dagger\hat{b}\\
 +\hbar g(\hat{a}^\dagger\hat{b}+\hat{a}\hat{b}^\dagger)+i\hbar\sqrt{\kappa}\epsilon_{c}(\hat{a}^\dagger-\hat{a})+i\hbar\sqrt{\kappa}\epsilon_p(\hat{a}^\dagger e^{-i\Omega t}
 -\hat{a} e^{i\Omega t}),
\end{aligned}
\end{equation}

where $G_{1}$ is the optomechanical coupling strength of mirror $M_1$ with the field in A and $G_2$ is the optomechanical coupling strength of mirrors $M_1$ and $M_2$ with the field in cavity B. Here $m_1$ and $m_2$ are the 'bare' masses of mirrors $M_1$ and $M_2$ respectively. We assume that the field in cavity B couples with equal strength to both mirrors $M_1$ and $M_2$ . Also $\hat{x}_{1}$, $\hat{x}_{2}$ and
$\hat{p}_{1}$, $\hat{p}_{2}$ are the position and momentum operators following the commutation relations, $[\hat{x}_{j},\hat{p}_{j}]=i\hbar$ ($j=1,2$) for mirrors $M_{1}$ and $M_{2}$.
The optomechanical coupling strengths are, $G_{1}=\frac{\omega_{1}}{L_{1}}$, $G_{2}=\frac{\omega_{2}}{L_{2}}$, where $\omega_{1}$($\omega_{2}$) are resonant frequencies
of cavities A and B respectively. Cavity A is driven by an external input laser field consisting of a strong control field and weak probe field denoted by 
$a_{in}(t)=\epsilon_{c}e^{-i\omega_{c}t}+\epsilon_p e^{-i\omega_p t}$ with field strengths $\epsilon_c$ and $\epsilon_p$ and frequencies $\omega_c$ and $\omega_p$ respectively.
The field strengths are given as $\epsilon_c=\sqrt{P_{c}/\hbar \omega_{c}}$ and $\epsilon_p=\sqrt{P_{p}/\hbar \omega_{p}}$ where $P_{c}$ and $P_{p}$ are the control and
probe field powers, respectively. Without loss of generality we assume that $\epsilon_c$ and $\epsilon_p$ are real. Here $\Delta_j=\omega_{c}-\omega_{j}$(j=1,2), is the cavity detuning of cavity A ($j=1$) and B ($j=2$)  while $\Omega=\omega_p-\omega_c$ is the detuning of the probe field with respect to the control field frequency $\omega_c$. The Hamiltonian in Eq.1 gives rise to the following quantum Langevin equation,

\begin{subequations}
\begin{equation} 
\frac{d\hat{a}}{dt}=i(\Delta_1+G_1\hat{x}_{1})\hat{a}-ig\hat{b}+\sqrt{\kappa}\epsilon_c+\sqrt{\kappa}\epsilon_pe^{-i\Omega t}-\frac{\kappa}{2}\hat{a}+\sqrt{\kappa}\hat{a}_{in},
\end{equation}
\begin{equation}
\frac{d\hat{b}}{dt}=i(\Delta_2+G_2[\hat{x}_{2}-\hat{x}_{1}])\hat{b}-ig\hat{a}-\frac{\kappa}{2}\hat{b}+\sqrt{\kappa}\hat{b}_{in},
\end{equation}
\begin{equation}
\frac{d\hat{x}_{j}}{dt}=\frac{\hat{p}_j}{m_j},
\end{equation}
\begin{equation}
\frac{d\hat{p}_1}{dt}=-m_1\Omega_1^2\hat{x}_1-\hbar G_2\hat{b}^\dagger\hat{b}+\hbar G_1\hat{a}^\dagger\hat{a}-\frac{\gamma_1}{2}\hat{p}_1+\sqrt{\gamma_1}\hat{\zeta}_{in},
\end{equation}
\begin{equation}
\frac{d\hat{p}_2}{dt}=-m_2\Omega_2^2\hat{x}_2+\hbar G_2\hat{b}^\dagger\hat{b}-\frac{\gamma_2}{2}\hat{p}_2+\sqrt{\gamma_2}\hat{\zeta}'_{in},
\end{equation}
\end{subequations}

where we have included the quantum and thermal noise operators following usual correlations,

\begin{subequations}\label{eq:3}
\begin{equation} 
<\hat{a}_{in}(t)\hat{a}^\dagger_{in}(t')>=\delta(t-t'),
\end{equation}
\begin{equation}
<\hat{b}_{in}(t)\hat{b}^\dagger_{in}(t')>=\delta(t-t'),
\end{equation}
\begin{equation}
<\hat{\zeta}_{in}(t)\hat{\zeta}^\dagger_{in}(t')>=\gamma_1(2\overline{n}_{th}+1)\delta(t-t'),
\end{equation}
\begin{equation}
<\hat{\zeta}'_{in}(t)\hat{\zeta}'^\dagger_{in}(t')>=\gamma_2(2\overline{n}'_{th}+1)\delta(t-t').
\end{equation}
\end{subequations}

Here $\hat{a}_{in}(t)$ and $\hat{b}_{in}(t)$ are the radiation shot noise destruction operators of fields in cavity A and B respectively. Also $\hat{\zeta}_{in}$ and $\hat{\zeta}_{in}^{'}(t)$ are mechanical noise destruction operators of mirrors $M_1$ and $M_2$ respectively. Also $\overline{n}_{th}$ and $\overline{n}'_{th}$ are the thermal occupation numbers for mirrors $M_1$ and $M_2$ respectively. We also assume that the strength of the control field is much larger than that of the probe field. 
Thus the average number of photons $\overline{n}_a$ and $\overline{n}_b$ are large, and hence the average displacements of the mirrors are large compared to their fluctuations. By equating the L.H.S of Eqs.~\ref{eq:2} to zero, we get the steady-state solutions, $\overline{a}, \overline{b}, \overline{x}_1, \overline{x}_2$ for operators $\hat{a},\hat{b},\hat{x_1},\hat{x_2}$ respectively, 

\begin{subequations}\label{eq:4}
\begin{equation} 
\overline{a}=\frac{-(i\overline{\Delta}_2-\kappa/2)\sqrt{\kappa}\epsilon_c}{(i\overline{\Delta}_1-\kappa/2)(i\overline{\Delta}_2-\kappa/2)+g^2},
\end{equation}
\begin{equation}
\overline{b}=\frac{-ig\sqrt{\kappa}\epsilon_c}{(i\overline{\Delta}_1-\kappa/2)(i\overline{\Delta}_2-\kappa/2)+g^2},
\end{equation}
\begin{equation}
\overline{x}_1=\frac{\hbar(G_1\mid\overline{a}\mid ^2-G_2\mid\overline{b}\mid ^2)}{m_1\Omega_1^2},
\end{equation}
\begin{equation}
\overline{x}_2=\frac{\hbar G_2\mid\overline{b}\mid^2}{m_2\Omega_2^2}.
\end{equation}
\end{subequations}

Here $\overline{\Delta}_{1}=\Delta_1+G_{1}\overline{x}_{1}$, $\overline{\Delta}_{2}=\Delta_2+G_{2}(\overline{x}_{2}-\overline{x}_{1})$. One implication in working with large control field is that the optomechanical coupling $G_{1,2}>\kappa$. This is the strong coupling regime, within which we can efficiently separate the optical and mechanical operators into their mean values and fluctuations. Here the probe field is treated as fluctuation(noise) since it is weak compared to the strength of the control field.  We also work within the resolved sideband regime, $\omega_j>>\kappa$, $\kappa$ being the optical decay rate.

\begin{figure}[ht]
\hspace{-0.7cm}
\begin{tabular}{cc}
\includegraphics [scale=0.85]{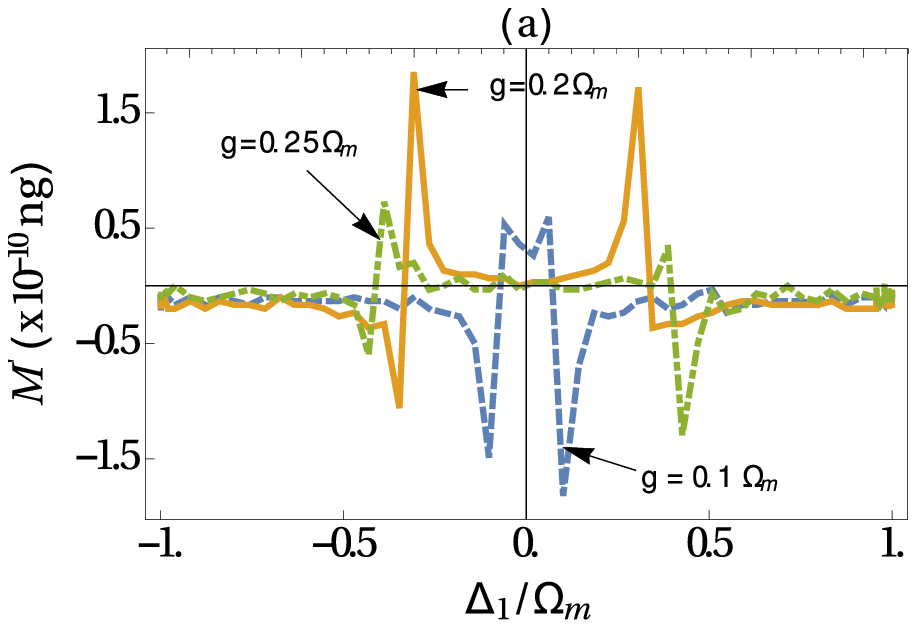} \includegraphics [scale=0.85] {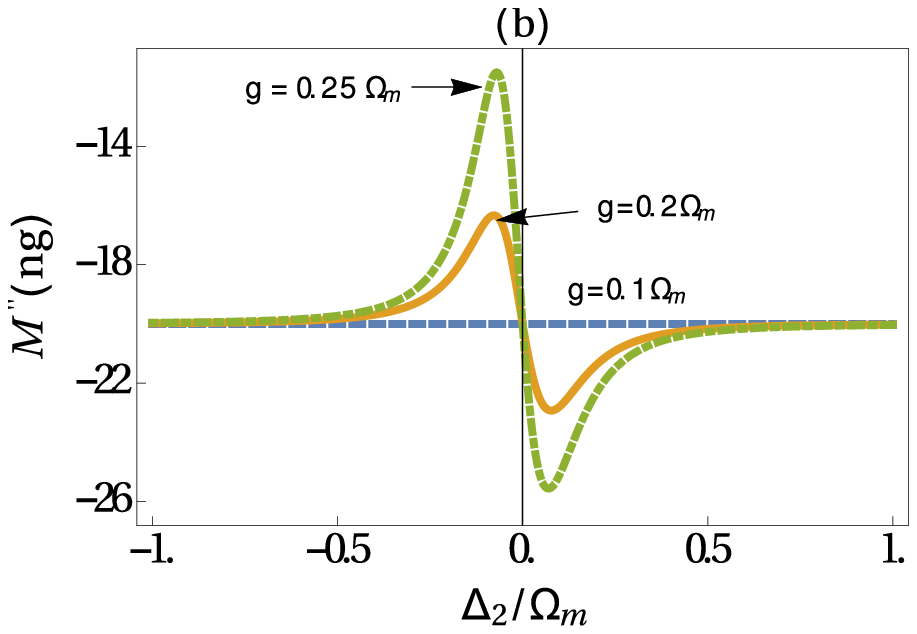}\\
 \end{tabular}
\caption{(Color online) Effective mass of mechanical resonators $M_{1}$, $M_{2}$ as a function of $\Delta_1/\Omega_m$(plot a) and  as function of $\Delta_2/\Omega_m$(plot b). The two mechanical resonator bare masses are
 $m_{1}=m_{2}=20ng$, $P_c=30mW$, the two optomechanical couplings $G_{1}=2\pi\times 18GHz/nm$, $G_{2}=2\pi\times 6GHz/nm$, $\gamma_{1}=\gamma_{2}=2\pi\times41kHz$, $\kappa=2\pi\times15MHz$, $\Delta_{2}=\Delta_1=\Omega_{m}$ and $\Omega_1=\Omega_2=\Omega_m=2\pi\times51.8MHz$.  } 
\label{fig:2}
\end{figure}

\section{Negative effective mass}
Negative masses are  not present in the real world but a composite system could give rise to an effective negative mass of components. The concept of effective negative mass is derived from condensed matter physics where an electron 'gains' an effective negative mass in the background lattice potential, in a way that its motion is in the opposite direction to the applied force. It has also been argued\citep{20} that effective mass in simple mass-spring systems can become negative in a certain frequency range of the driving force. Here, we look at the behaviour of the system after it reaches a steady state, under the influence of a constant restoring/equilibrium force. 

We calculate the equilibrium forces $F_{j}$(j=1,2) on the mechanical mirrors $M_{1}$ and $M_{2}$ such that $F_j=-\frac{\partial H}{\partial x_j}$,
H being the Hamiltonian given in Eq.~\ref{eq:1}. The initial equilibrium values i.e the initial positions of the mirrors can be offset. 

The force equation for the $j^{th}$ resonator(j=1,2) is,\\

\begin{subequations}\label{eq:5}
\begin{equation} 
  F_{j}=-\frac{\partial H}{\partial x_{j}},
\end{equation}
\begin{equation}
 F_1=\hbar G_{1}<\hat{a}^{\dagger}\hat{a}>-\hbar G_{2}<\hat{b}^{\dagger}\hat{b}>-m_{1}\Omega_{1}^2\overline{x}_{1},
\end{equation}
\begin{equation}
 F_2=\hbar G_{2}<\hat{b}^{\dagger}\hat{b}>-m_{2}\Omega_{2}^2\overline{x}_{2}.
 \end{equation}
 \end{subequations}

Substituting the steady state values $\overline{x}_1$ and $\overline{x}_2$ from Eq.~\ref{eq:4} and considering large classical fields we get,\\

\begin{subequations}
\begin{equation} \label{eq:6}
  |\overline{a}|^2=\frac{g^2|\overline{b}|^2+\kappa\epsilon_{c}^2}{(\overline{\Delta}_{1}^2+\kappa^2/4)},\ \ \ \ \
  |\overline{b}|^2=\frac{g^2|\overline{a}|^2}{(\overline{\Delta}_{2}^2+\kappa^2/4)},
  \end{equation}
  \begin{equation}
  |\overline{a}|^2\approx \frac{g^2|\overline{b}|^2+\kappa\epsilon_{c}^2}{\left(\Delta_{1}^2+\kappa^2/4\right)}\left(1+\frac{2\Delta_{1}G_{1}\overline{x}_{1}}{(\Delta_{1}^2+\kappa^2/4)}\right),
 \end{equation}
 \begin{equation}
 |\overline{b}|^2\approx\frac{g^2|\overline{a}|^2}{(\Delta_{2}^2+\kappa^2/4)}\left(1+\frac{2\Delta_{2}G_{2}(\overline{x}_{1}-\overline{x}_{2})}{(\Delta_{2}^2+\kappa^2/4)}\right).
\end{equation}
\end{subequations}

Since $\overline{\Delta}_{1}=\Delta_{1}+G_{1}\overline{x}_{1}$
and $\overline{\Delta}_{2}=\Delta_{2}+G_{2}(\overline{x}_{2}-\overline{x}_{1})$, we perform a Taylor expansion upto first order under the assumption, $\frac{\overline{x}_{1}}{L_{1}}<<1$ and $\frac{(\overline{x}_{2}-\overline{x}_{1})}{L_{2}}<<1$  to obtain Eqns.(6b) and (6c) from Eqn.(6a). Solving these equations, we get the equilibrium forces on resonators $M_{1}$ and $M_{2}$ as,

\begin{subequations}\label{eq:7}
\begin{equation}
  F_{1}=M'\Omega_{1}^2\overline{x}_{1},
\end{equation}
\begin{equation}
  F_{2}=M''\Omega_{2}^2\overline{x}_{2}.
\end{equation}
\end{subequations}

where $M'$ and $M''$ are the effective masses of resonators $M_{1}, M_{2}$ respectively. The expressions for $M'$ and $M''$ are given explicitly in Appendix A.
From Fig.~\ref{fig:2} above we see that the effective mass of mirror $M_1$, as function of $\Delta_1/\Omega_m$, is almost zero as compared to that of mirror $M_2$. Note that for all future analysis we have taken $\Omega_1=\Omega_2=\Omega_m$. On the other hand the effective mass of mirror $M_2$ shows a significant variation around  $\Delta_2=0$ and approaches a constant value of -20 ng for  $|\Delta_2|>0.5 \Omega_{m}$.  For photon tunneling $g=0.1 \Omega_{m}$, the effective mass of mirror $M_2$ shows a constant value of -20 ng. Sharp and significant variation in  $M''$  is noticed as the tunneling increases beyond  $g=0.1 \Omega_{m}$ clearly demonstrating that the effective mass can be tuned with the tunneling parameter.  It has been shown that quantum back-action in cavity optomechanics can be avoided using negative-mass BECs\citep{32} and negative mass reference frame of spin-oscillator (caesium atom ensemble in magnetic field) \citep{33}. One might ask if it is possible to use negative effective mass of mechanical mirrors to achieve a similar effect.

\section{Controllable Multistability}
Multistability of the displacement of a mechanical oscillator coupled to a cavity field, is an indication of the optical spring effect due to optomechanical interaction between the cavity field and the oscillator. For non zero detuning, multistability is an indication of the dynamical back-action and leads to the heating or cooling of the mechanical oscillator. This is because a multistable behaviour implies a change in decay rate and the resonance frequency of the oscillator. Multistability is also an essential ingredient when it comes to the designing of all optical switches\citep{24}. It is an obvious advantage if the multistable behavior occurs at low values of the input power.\\ In this section we show that multistability for the displacement of mechanical mirror $M_1$ can be modified and controlled by varying the tunnelling rate g. Using Eqs.~\ref{eq:4} we get a polynomial for the displacement of mirror $M_1$ ($\overline{x}_{1}$). For further simplification we assume that $\Theta_1=\Theta_2$ ($\Theta_j=i\overline{\Delta}_j-\kappa/2$, $j=1,2$). This gives rise to a fifth-order polynomial in $\overline{x}_{1}$. From Eq. 4, and using $\Theta_1=\Theta_2$ we get,

\begin{subequations}\label{eq:8}
    \begin{equation}
    |\overline{a}|^2=\frac{(-i\overline{\Delta_1}+\frac{\kappa}{2})(i\overline{\Delta_1}+\frac{\kappa}{2})\kappa\epsilon_c^2}{((i\overline{\Delta_1}-\frac{\kappa}{2})^2+g^2)((i\overline{\Delta_1}+\frac{\kappa}{2})^2+g^2)},
    \end{equation}
    \begin{equation}
         |\overline{b}|^2=\frac{g^2\kappa\epsilon_c^2}{((i\overline{\Delta_1}-\frac{\kappa}{2})^2+g^2)((i\overline{\Delta_1}+\frac{\kappa}{2})^2+g^2)}.
\end{equation}
\end{subequations}\\

Substituting the above equations in Eq.4c gives,

\begin{equation} \label{eq:9}
\begin{aligned}
a_{1}\overline{x}_{1}^5+a_{2}\overline{x}_{1}^4+a_{3}\overline{x}_{1}^3+a_{4}\overline{x}_{1}^2+a_{5}\overline{x}_{1}+a_{6}=0,
\end{aligned}
\end{equation}

where the coefficients are,

\begin{subequations}\label{eq:10}
\begin{equation} 
a_{1}=G_{1}^4,
\end{equation}
\begin{equation}
a_{2}=4{\Delta_{1}}{G_{1}}^3,
\end{equation}
\begin{equation}
a_{3}=-2 g^2{G_{1}}^2 + 6{\Delta_{1}}^2{G_{1}}^2 + \frac {{G_{1}}^2{\kappa^2}}{2},
\end{equation}
\begin{equation}
a_{4}=-4{\Delta_{1}}g^2{G_{1}} - \frac {{G_{1}}^3 \kappa\epsilon_c^2\hbar} {{m_{1}}{\Omega_{1}}^2} + 4{\Delta_{1}}^3{G_{1}} + {\Delta_{1}}{G_{1}}{\kappa^2},
\end{equation}
\begin{equation}
a_{5}={\Delta_{1}}^4 + \frac {{\Delta_{1}}^2{\kappa^2}}{2} + g^4 - 2{\Delta_{1}}^2 g^2 + \frac {g^2{\kappa^2}}{2} - \frac {2 {\Delta_{1}}{G_{1}}^2 \kappa\epsilon_c^2\hbar} {{m_{1}}{\Omega_{1}}^2} + \frac {{\kappa^4} }{16},
\end{equation}
\begin{equation}
a_{6}=\frac {g^2{G_{2}}\kappa\epsilon_c^2\hbar} { {m_{1}}{\Omega_{1}}^2} - \frac {{\Delta_{1}}^2{G_{1}}\kappa\epsilon_c^2\hbar} {{m_{1}}{\Omega_{1}}^2} - \frac {{G_{1}}{\kappa^3\epsilon_c^2\hbar} }{4{m_{1}}{\Omega_{1}}^2}.
\end{equation}
\end{subequations}

\begin{figure}[ht]
\hspace{-1.3cm}
\begin{tabular}{cc}
\includegraphics [scale=0.70]{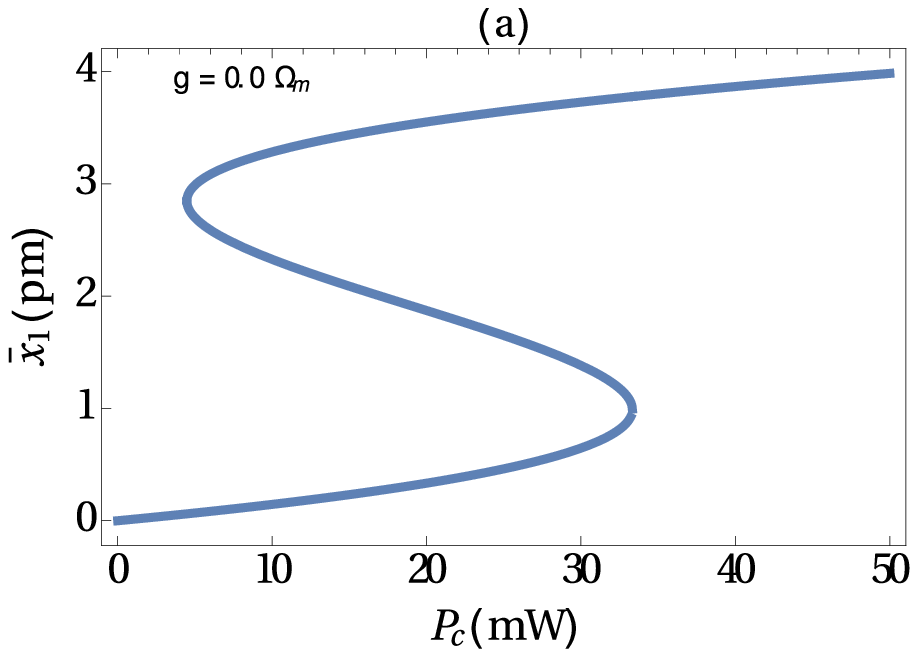} \hspace{4mm} \includegraphics [scale=0.70] {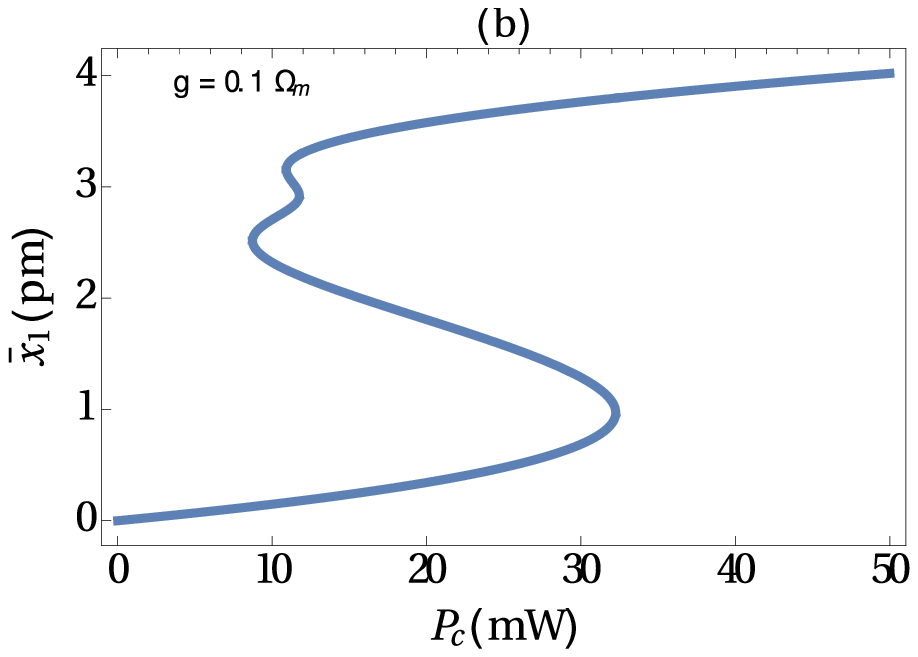}\\
\includegraphics [scale=0.70]{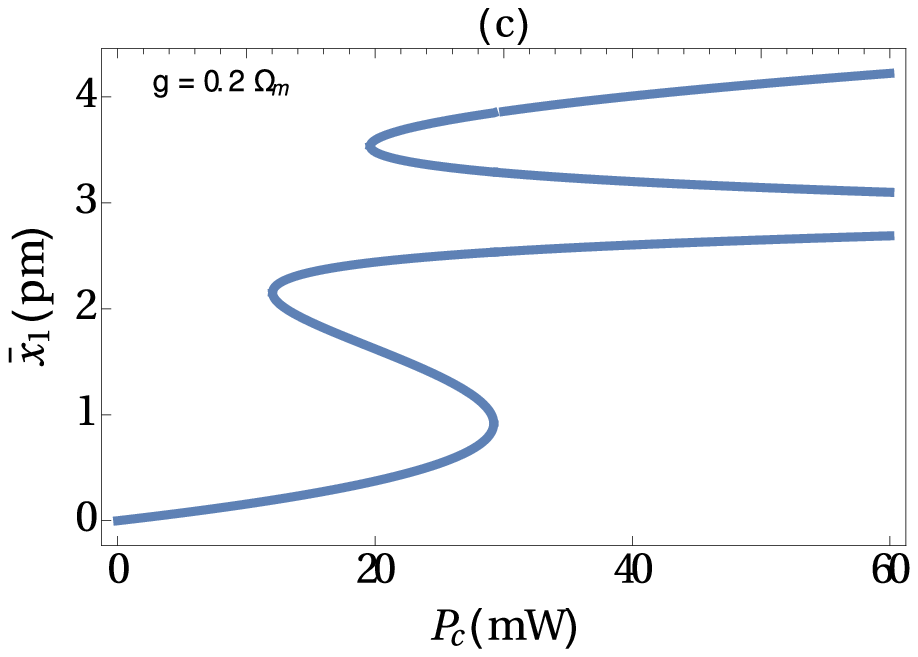} \hspace{4mm} \includegraphics [scale=0.70] {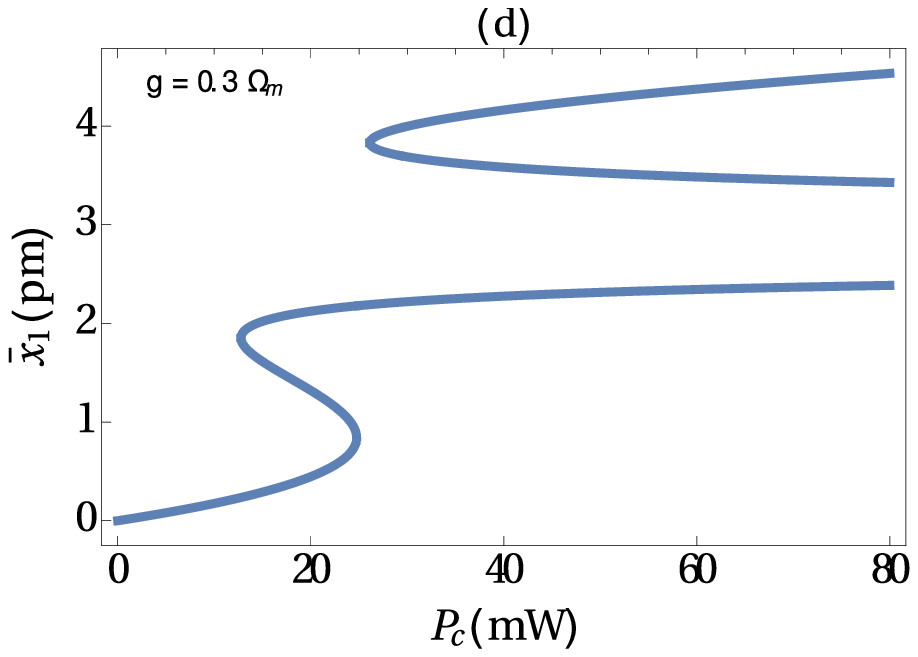}\\
\includegraphics [scale=0.70]{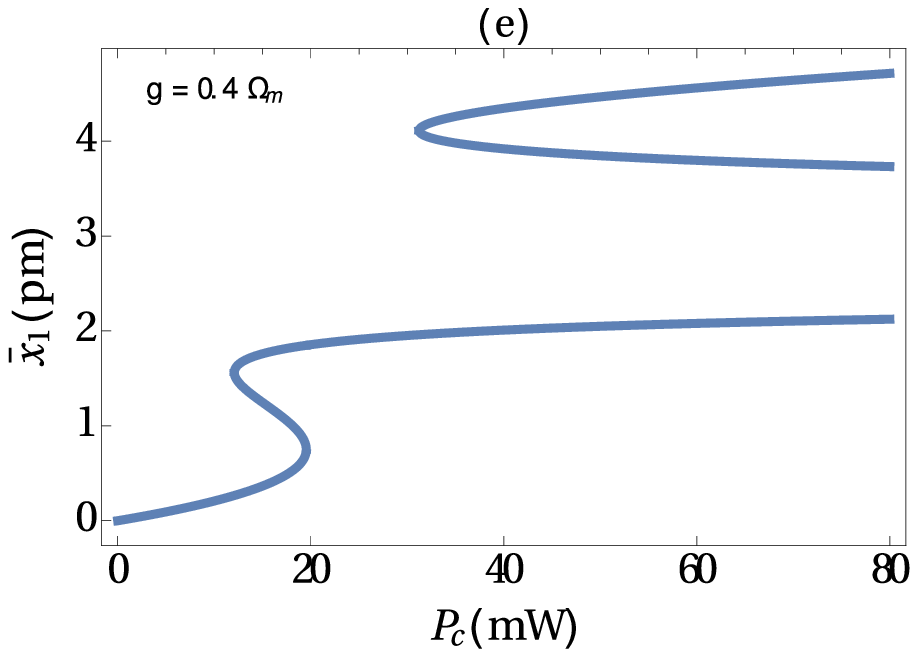} \hspace{4mm} \includegraphics [scale=0.70] {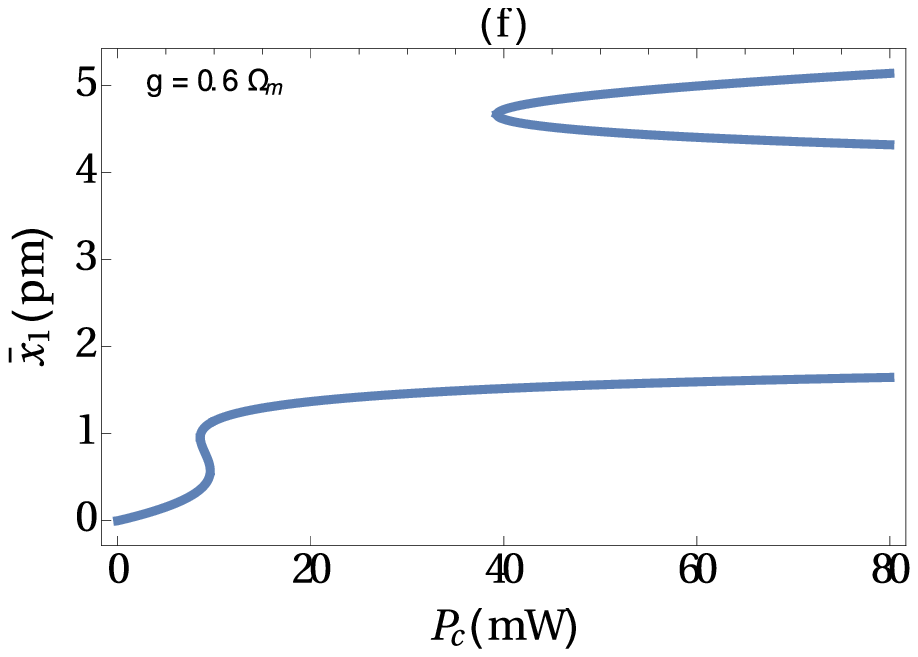}\\
\end{tabular}
\caption{(Color online)The displacement of the middle mechanical mirror $\overline{x}_{1}$ as a function of the power of the control field $P_{c}$(in mW) for different values of tunnelling rate g.
For calculations we use experimental parameters from ref.\citep{8};The parameters used are; $m_{1}=m_{2}=20ng$, the two optomechanical couplings $G_{1}=2\pi\times18GHz/nm$,  $G_{2}=2\pi\times6GHz/nm$, $\gamma_{1}=\gamma_{2}=2\pi\times41kHz$, $\kappa=2\pi\times15MHz$, $\Delta_{1}=-\Omega_{m}$  and $\Omega_1=\Omega_2=\Omega_m=2\pi\times51.8MHz$.}
\label{fig:3}
\end{figure}

 In Fig.3 we have plotted the stationary values of the displacement of mirror $M_{1}$ as a function of the power of the control field $P_{c}$ for different values of the tunneling rate 'g'. It is interesting to see that mechanical mirrors coupled via cavity fields tend to produce multistable behavior for the steady-state values of the displacement of $M_{1}$.\\ As seen in the fig.~\ref{fig:3}, the system goes from being bistable to being multistable by changing the tunnelling rate of the middle mirror $M_{1}$.    For $g=0$ the steady-state value $\overline{x}_{1}$ shows a bistable behavior infered from the  S-shape in Fig.3(a). This is because at $g=0$, only the mode in cavity A couples with the middle mirror $M_{1}$.

\begin{figure}[ht]
\hspace{-1.3cm}
\begin{tabular}{cc}
\includegraphics [scale=0.70]{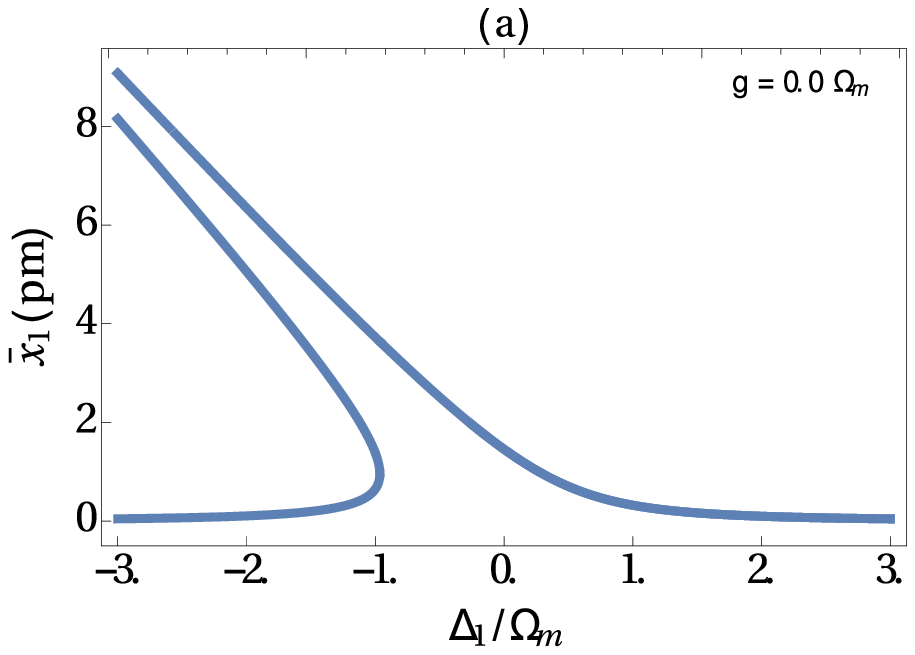} \hspace{4mm} \includegraphics [scale=0.70] {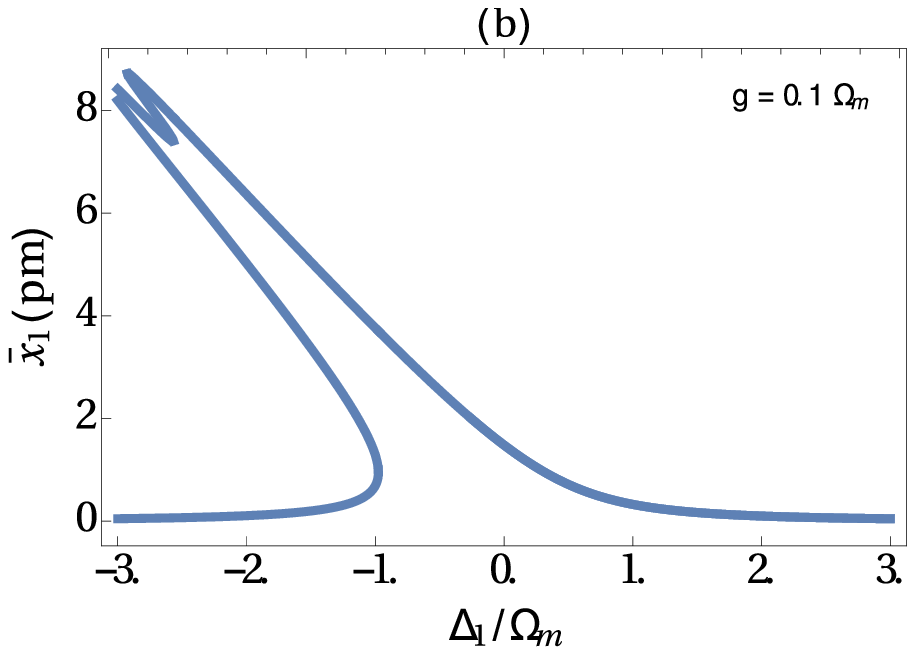}\\
\includegraphics [scale=0.70]{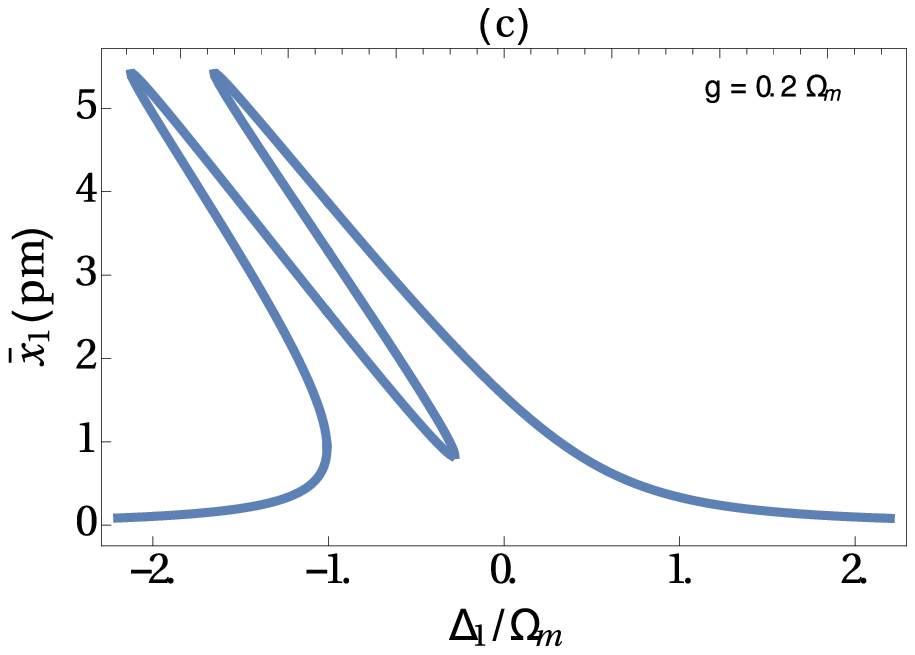} \hspace{4mm} \includegraphics [scale=0.70] {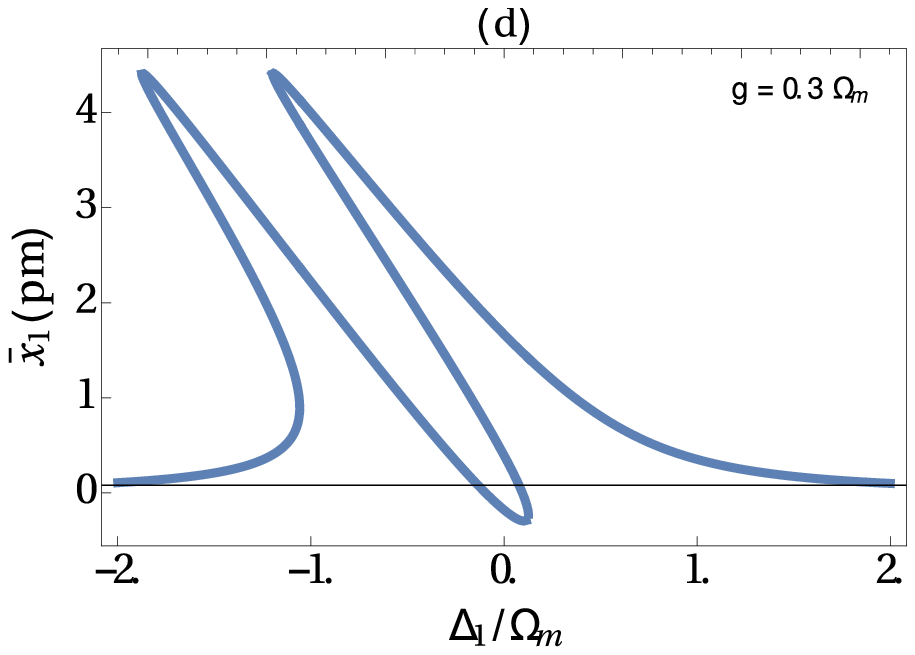}\\
\includegraphics [scale=0.70]{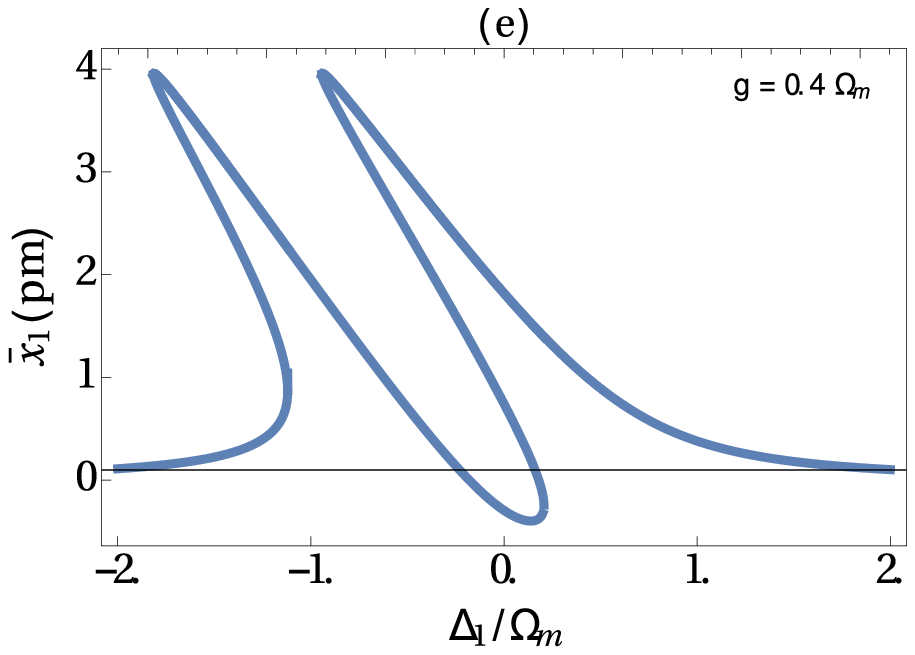} \hspace{4mm} \includegraphics [scale=0.70] {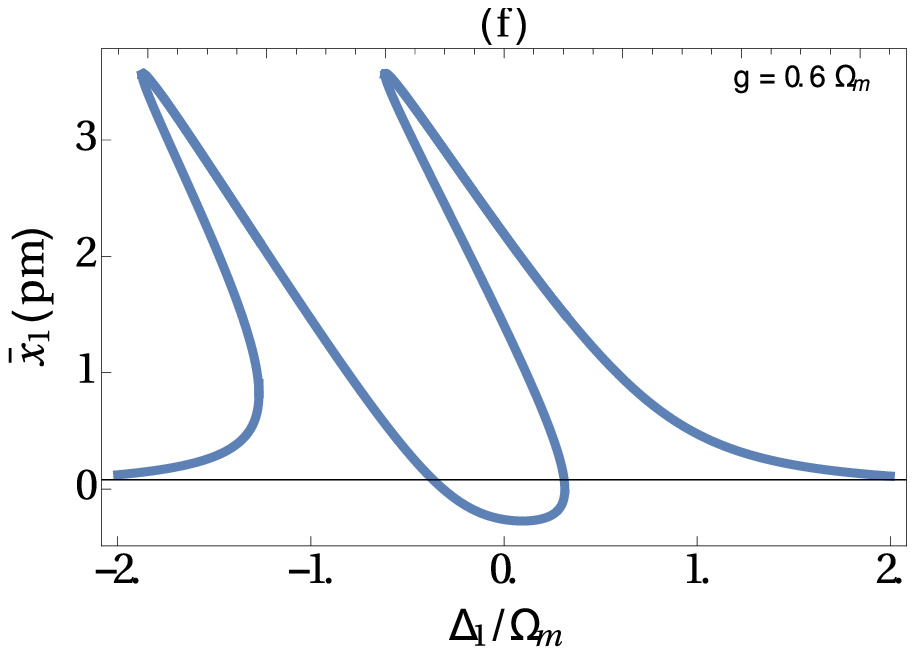}\\
\end{tabular}
\caption{(Color online)The displacement of the middle mechanical resonator $\overline{x}_{1}$ as a function of detuning $\Delta_{1}$ of field in cavity A for different values of tunnelling
rate g.The parameters used are; $m_{1}=m_{2}=20ng$, the two optomechanical couplings $G_{1}=2\pi\times18GHz/nm$, $G_{2}=2\pi\times6GHz/nm$,$\gamma_{1}=\gamma_{2}=2\pi\times41kHz$, $\kappa=2\pi\times15MHz$, $P_c=30mW$  and $\Omega_1=\Omega_2=\Omega_m=2\pi\times51.8MHz$.}
\label{fig:4}
\end{figure}

At higher tunneling rates $(g=0.2\Omega_m,0.4\Omega_{m},..., 0.6\Omega_m)$, cavity B gets activated and the coupling of mirrors $M_1$ and $M_2$ with the field of cavity B in addition to that of cavity A gives rise to a multistable behaviour for the displacement of $M_1$, as shown in Fig.~\ref{fig:3}(b-f). As one increases the tunneling rate g, the multistable behaviour of $M_1$ shifts towards higher values of the input power. At lower values of the tunnelling rate ($g\approx 0.1\Omega_m$), the system gives rise to five stable points. It is noteworthy that this happens at low values of the input power($P_c=30mW$). We thus have a system which can provide  bistable and multistable behaviors at low values of the input power by simply controlling the photon tunneling rate between the two cavities. Besides its use in "all optical switches", it can also be used to build logic gate and memory devices for quantum information processing where low energy input is a necessity. Experimentally, the controlled multi-state switching can be realize by additing a pulse sequence into the input field \citep{sheng}.
In fig.~\ref{fig:4}, we plot the displacement of mirror $M_{1}$ with respect to the detuning $\Delta_{1}$ of field in cavity A. We see that even in the absence of tunnelling between the two cavities, the plot of displacement $x_{1}$ indicates non-linear behaviour in the red-detuned region whereas the stable region appears when the field in cavity A is blue-detuned. It is noteworthy that Fig.4 also displays the appearance of two bistable points as opposed to a single bistability observed for $g=0$ (Fig. 4(a)). As $g$ increases from $0.1 \Omega_{m}$ to $0.6 \Omega_{m}$ one of the bistable point shifts from the red-detuned region to the blue-detuned side.

\begin{figure}[h]
 \hspace{-1.3cm}
  \begin{tabular}{c}
  \includegraphics[scale=0.70]{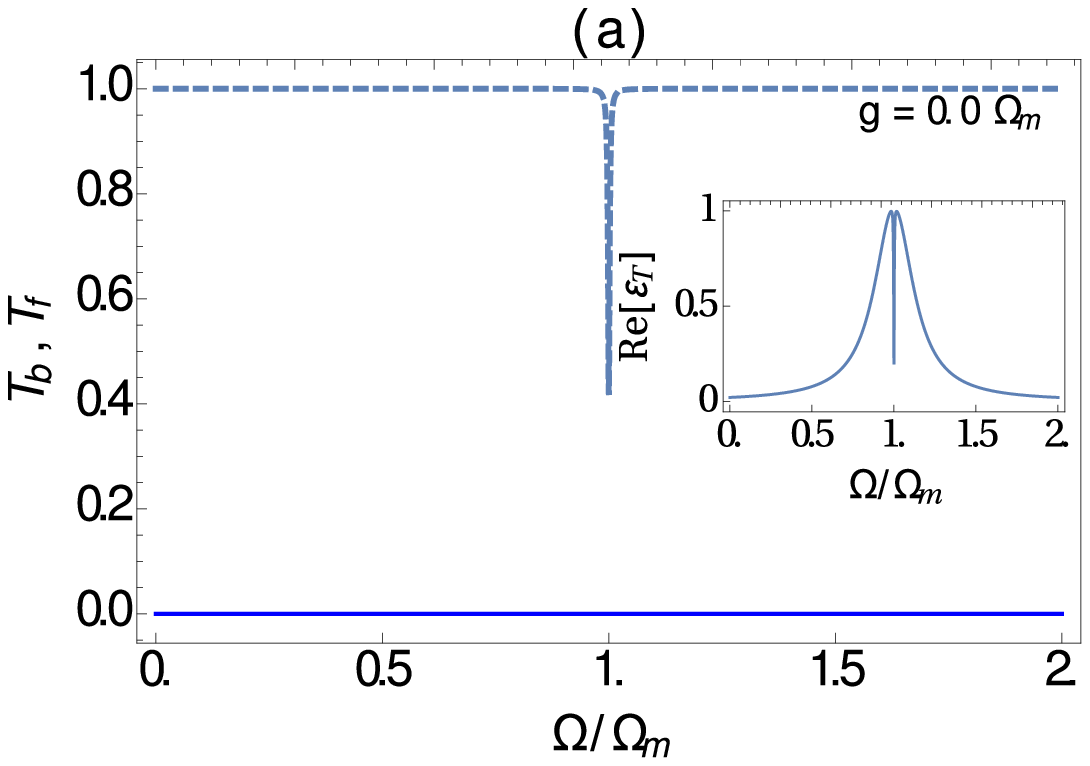}\\
  \includegraphics[scale=0.85]{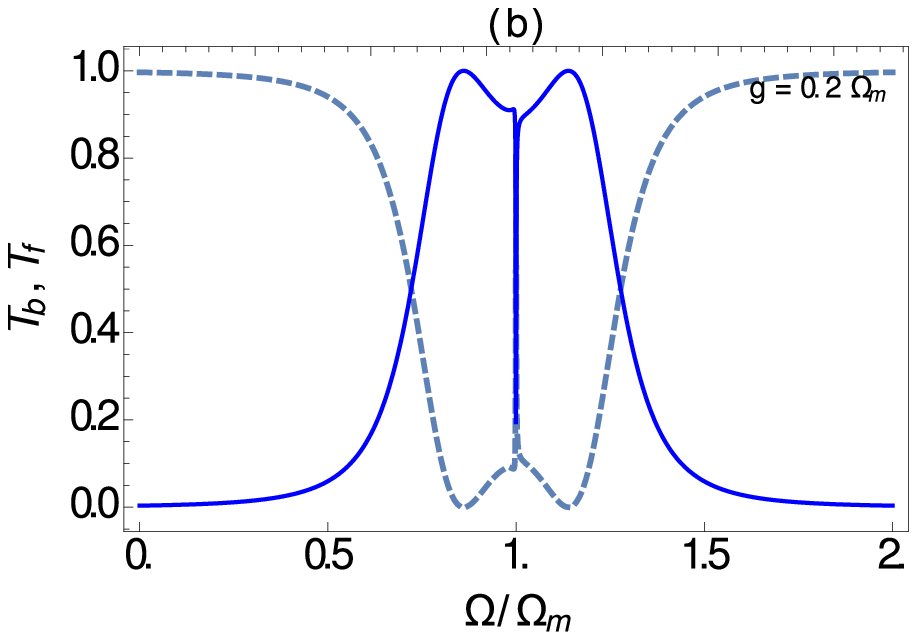}\\
  \includegraphics[scale=0.85]{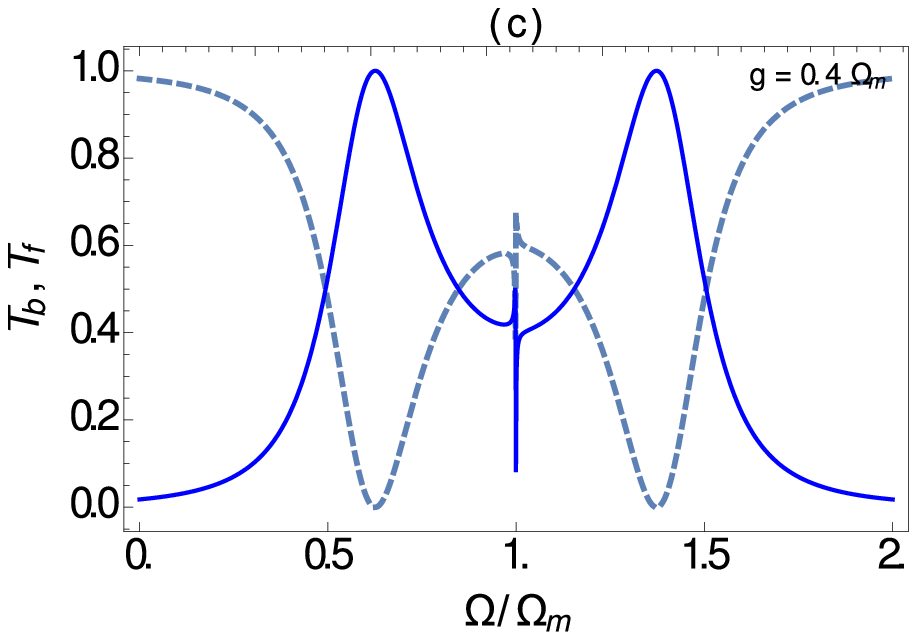}\\
  \end{tabular}
\caption{Normalized forward power transmission $T_{f}$ and backward reflection coefficients $T_{b}$ as a function of the probe detuning $\Omega$ for different values of the tunnelling rate g.The dashed line is the backward reflection coefficient $T_b$ while the solid line is the forward transmission coefficient $T_f$. The inset in Fig. 5a depicts real value of the response function $\epsilon_T$.The parameters used are; $P_{c}=1mW$, $m_{1}=m_{2}=20ng$, $G_{1}=G_{2}=2\pi\times6 GHz/nm$,$\gamma_{1}=\gamma_{2}=2\pi\times41kHz$, $\kappa=2\pi\times15MHz$,
$\Delta_{1}=\Delta_{2}=-\Omega_{m}$  and $\Omega_1=\Omega_2=\Omega_m=2\pi\times51.8MHz$.}
\label{fig:5}
\end{figure}

\section{Controllable OMIA and Fano resonance}
Analogous to the phenomenon of Electromagnetically Induced Transparency(EIT) demonstrated in atomic ensembles, OptoMechanically Induced Transparency(OMIT) is seen in optomechanical
systems \citep{20}. The presence of a strong control field and weak probe field induces a beat frequency on the mechanical resonator. Optomechanical interaction creates dressed phonon-photon states. This is similar to atom-photon dressed states created due to the AC Stark (Autler-Townes) effect in EIT. The red-detuning of the control laser leads to a destructive interference between photons excited through different pathways. This gives rise to a transparency window in the intracavity field. The weak probe field, as the name suggests, 'probes' this transparency.\\
 Since we consider a single port/reflection cavity, wherein the output is measured on the same port as the input, the expression for backward reflection coefficient $T_{b}$ shows OMIA (OptoMechanically Induced Absorption \citep{34}) rather than OMIT. If a WGM microresonator is used instead of a single input output Fabry-Perot cavity, we are able to see OMIT \citep{15a}.  In Fig.~\ref{fig:5} we plot the normalized forward power transmission $T_{f}$ and backward reflection $T_{f}$ as a function of the probe detuning $\Omega$(in units of $\Omega_{m}$) under the three different values of the photon tunneling parameter $g$ based on the obtained analytical expressions ($B6a$) and ($B6b$) of Appendix B  for red-detuned cavity fields, i.e. $\Delta_{1}=\Delta_{2}=-\Omega_{m}$. There is no characteristic change in the absorption spectrum defined as $Re[\epsilon_T]$, where $\epsilon_T=\sqrt{\kappa}A_1^-/\epsilon_p$. It shows a dip in the absorption at $\Omega=\Omega_m$ (inset of Fig.~\ref{fig:5}a) as expected. The expression for $A_{1}^{-}$ is derived in Appendix B. At zero tunnelling rate($g=0$),

\begin{equation}
    \begin{aligned}
    \epsilon_T=\kappa\left(-\frac{1}{D_1}-\frac{i |\overline{a}|^2 G_1}{D_1^2(-\frac{i|\overline{a}|^2 G_1}{D_1}-\frac{1}{\hbar \chi_1 G_1}+\frac{i |\overline{a}|^2 G_1}{D_2^*})}\right),\\
    \end{aligned}
\end{equation}

where $D_1=i(\Omega+\overline{\Delta}_1)-\frac{\kappa}{2}$ and $D_2^*=i(\Omega-\overline{\Delta}_1)-\frac{\kappa}{2}$. This gives,

\begin{equation}
    \begin{aligned}
    \epsilon_T=\kappa\left(\frac{(\Omega^2-\Omega_m^2+i\gamma_m\Omega)(\frac{\kappa}{2}-i(\Omega-\overline{\Delta}_1)-2i\beta\Omega_m}{(\Omega^2-\Omega_m^2+i\gamma_m\Omega)((\frac{\kappa}{2}-i\Omega)^2+\overline{\Delta}_1^2)-4\overline{\Delta}_1\beta\omega_m}\right),
    \end{aligned}
\end{equation} 

where $\beta=|\overline{a}|^2G_1^2x_{zpf}^2$ and $x_{zpf}=\sqrt{\frac{\hbar}{2 m \omega_m}}$. We substitute $y=\Omega-\Omega_m$, $\overline{\Delta}_1\sim-\Omega_m$ and after some simplifications we get,

\begin{equation}
    \begin{aligned}
    \epsilon_T=\frac{\kappa}{\frac{\kappa}{2}-iy+\frac{\beta}{\frac{\gamma_m}{2}-iy}}.
    \end{aligned}
\end{equation}. 

The case of $g=0$ means that the control field entering cavity A simply reflects back from mirror $M_{1}$ which leads to $T_{f}=0$ as seen in Fig. 5(a). At $g=0$, the mode in cavity B is not introduced. This corresponds to the same situation encountered in Fig. 3(a) and 4(a) where the optomechanical system possess only one single bistablity.  The Fano resonance, which has a distinct sharp asymmetric line-shape profile is considerably different from OMIT and OMIA spectral profile shown in Fig. 5(a). Due to the sharp asymmetric line-shape shown in Fig. 5(b) and 5(c), small changes in the probe detuning $\Omega$ is able to cause significant change in $T_{b}$ and $T_{f}$. This is distinctly visible around $\Omega=\Omega_{m}$. Consequently, the possibility to coherently control and tune the Fano resonance is extremely important.
The physical mechanism for the generation of such Fano resonance is the destructive interference between the forward and backward reflected optical fields introduced via the coupling term $\hbar g (\hat{a}^{\dagger} \hat{b}+\hat{a} \hat{b}^{\dagger})$. By comparing Fig. 5(b) and 5(c), Fano line-shapes changes distinctly with the increase of the tunneling rate from $g=0.2 \Omega_{m}$ to $g=0.4 \Omega_{m}$. In particular, we notice that the dip at $\Omega=\Omega_{m}$ of the Fano resonance in $T_{f}$ is relatively small when $g=0.4 \Omega_{m}$ compared to the case when $g=0.2 \Omega_{m}$. The forward transmission contrast of the Fano response in our optomechanical system is approximately $70 \%$ (see Fig. 5(b)), which makes the system suitable for Telecom devices \citep{reed}. Hence, we are able to effectively control and tune the Fano line-shapes by approximately changing the photon tunneling rate $g$ between the two optical cavities A and B. It is also possible to measure the forward transmission coefficient $T_{f}$ using an output port at mirror $M_{2}$. The detailed derivation of $T_{b}$ and $T_{f}$ are explicitely given in Appendix B.

\section{Conclusion}
We have proposed a novel scheme for generating and controlling negative effective mass, optical multistability and sharp asymmetric Fano resonances in a double cavity optomechanical setup with one stationary and two harmonically bound mirrors. We demonstrate that under appropriate conditions, the effective mass of one of the movable mirrors can exhibit negative values which can be controlled and tuned by the photon tunneling parameter and the cavity detuning. This result may provide new insights into the aspects of designing optomechanical metamaterials. When $g=0$, there is only one bistable region. Interestingly, the bistability can turn into multistability in the beginning and then become two separate bistable regions with increasing $g$. The optomechanical bistability and multistability occurs at low values of the input power and can be controlled by the photon tunneling rate between the two cavities. With realistic experimental parameters, we observe the sharp asymmetric Fano resonance line-shapes. The Fano spectral profile can again be tuned by appropriately changing $g$. Our results demonstrate that the present scheme will be helpful in practical applications, such as optical switches and optical sensors.

\section{Acknowlegement}
A. B. B acknowledges Birla Institute of Technology and Science , Pilani for the facilities to carry out this research. V.N.P would like to thank the School of Physical Sciences, JNU for their support and would also like to thank  group members, Deepti Sharma and Shahnoor Ali for their valuable inputs.

\appendix
\section{}
The expressions for the effective negative mass of mirrors $M_1$ and $M_2$ are given as,

\begin{subequations}
\begin{equation}
 \begin{aligned}
M'=-m_1+\frac{2 \Delta_2 \zeta_1 g^2 G_2^2 (1-\Lambda_1) \hbar }{\Lambda_2 \Omega_1^2 \left(\Delta_2^2+\frac{\kappa ^2}{4}\right) \left(\Delta_2^2+\frac{\kappa^2}{4}\right)}
   -\frac{2\Delta_2 \zeta_1 g^2 G_2^2 \hbar }{\Omega_1^2 \left(\Delta_2^2+\frac{\kappa ^2}{4}\right) \left(\Delta_2^2+\frac{\kappa^2}{4}\right)}-
   \frac{\zeta_2 g^2 G_2 \hbar }{\Omega_1^2 \left(\Delta_2^2+\frac{\kappa^2}{4}\right)}\\+
   \frac{\zeta_3 g^2 G_2 (1-\Lambda_1) \hbar }{\Lambda_2 \Omega_1^2 \left(\Delta_2^2+\frac{\kappa^2}{4}\right)}+
   \frac{\zeta_2 G_1 \hbar }{\Omega_1^2}-\frac{\zeta_3 G_1 (1-\Lambda_1) \hbar}{\Lambda_2\Omega_1^2},\\\\
\end{aligned}
\end{equation}
\begin{equation}
\begin{aligned}
M''=-m_2+\frac{2\Delta_2 \zeta_1 g^2 G_2^2 \Lambda_2\hbar}{(1-\Lambda_1) \Omega_2^2 \left(\Delta_2^2+\frac{\kappa^2}{4}\right) \left(\Delta_2^2+\frac{\kappa^2}{4}\right)}
    -\frac{2 \Delta_2 \zeta_1 g^2 G_2^2 \hbar }{\Omega_2^2 \left(\Delta_2^2+\frac{\kappa ^2}{4}\right)\left(\Delta_2^2+\frac{\kappa_2^2}{4}\right)}\\
    +\frac{{\zeta_2} g^2{G_2} {\Lambda_2} \hbar }{(1-{\Lambda_1}){\Omega_2}^2 \left({\Delta_2}^2+\frac{{\kappa}^2}{4}\right)}
    -\frac{{\zeta_3} g^2{G_2} \hbar }{{\Omega_2}^2\left({\Delta_2}^2+\frac{{\kappa}^2}{4}\right)},\\\\
 \end{aligned}
\end{equation}
\end{subequations}

where,

\begin{subequations}
\begin{equation}
\begin{aligned}
A_1=\frac{g^2}{\Delta_1^2+\frac{\kappa^2}{4}}, 
A_2=\frac{g^2}{\Delta_2^2+\frac{\kappa^2}{4}}, 
B_1=\frac{\Omega_l^2}{\Delta_1^2+\frac{\kappa^2}{4}}, 
B_2=\frac{2\Delta_2 G_2}{\Delta_2^2+\frac{\kappa^2}{4}},
\end{aligned}
\end{equation}
\begin{equation}
\begin{aligned}
C_1=-\frac{2 \Delta_1 G_1}{\Delta_1^2+\frac{\kappa^2}{4}}, 
\zeta_1=\frac{B_1}{1-A_1 A_2}, 
\zeta_2=\frac{B_1 (A_1 A_2 B_2+C_1)}{(1-A_1 A_2)^2}, 
\zeta_3=\frac{A_1 A_2 B_1 B_2}{(1-A_1 A_2)^2},
\end{aligned}
\end{equation}
\begin{equation}
 \begin{aligned}
\Lambda_1=\frac{\zeta_2 G_1 \hbar }{m_1 \Omega_1^2}-\frac{G_2 \psi_2\hbar }{m_1\Omega_1^2}, 
\Lambda_2=\frac{G_2 \psi_3 \hbar }{m_1 \Omega_1^2}-\frac{\zeta_3 G_1 \hbar }{m_1 \Omega_1^2},
\end{aligned}
\end{equation}
\begin{equation}
\begin{aligned}
\psi_1=A_2 \zeta_1, \psi_2=A_2(\zeta_2+B_2 \zeta_1),\psi_3=A_2(\zeta_3+B_2\zeta_1).\\
\end{aligned}
\end{equation}
\end{subequations}

\section{}
Using a mean field approximation, Eqs.2 can be  written in terms of their steady-state values and fluctuations. The fluctuating terms in Eqs.~\ref{eq:2} can be written as,

\begin{subequations}
\begin{equation}
\begin{aligned}
\frac{d}{dt}\delta a&=\Theta_1\delta{a}+iG_1\delta{x}_1(\overline{a}+\delta{a})
-ig\delta{b}+\sqrt{\kappa}\epsilon e^{-i\Omega t},\\\\
\end{aligned}
\end{equation}
\begin{equation}
\begin{aligned}
\frac{d}{dt}\delta b&=\Theta_2\delta{b}+iG_2(\delta{x}_2-\delta{x}_1)(\overline{b}+\delta{b})
-ig\delta{a},\\\\
\end{aligned}
\end{equation}
\begin{equation}
\begin{aligned}
 \Psi_{1}\delta{x}_1&=\frac{-\hbar G_2}{m_1}(\overline{b}^*\delta{b}+\overline{b}\delta{b}^*)
 +\frac{\hbar G_1}{m_1}(\overline{a}^*\delta{a}+\overline{a}\delta{a}^*),\\\\
\end{aligned}
\end{equation}
\begin{equation}
\begin{aligned}
 \Psi_{2}\delta{x}_2&=\frac{\hbar G_2}{m_2}(\overline{b}^*\delta{b}+\overline{b}\delta{b}^*),
\end{aligned}
\end{equation}
\end{subequations}

where,

$\Theta_j=i\overline{\Delta}_j-\kappa_j/2$, $\overline{\Delta}_{1}=\Delta_1+G_{1}\overline{x}_{1}$, $\overline{\Delta}_{2}=\Delta_2+G_{2}(\overline{x}_{2}-\overline{x}_{1})$
and $\Psi_{j}=\frac{d^2}{dt^2}+\omega_{j}^2+m_{j}\frac{\gamma_{j}}{2}\frac{d}{dt}$\\
Since the fluctuations are small compared to their corresponding  mean field values , in calculating the equations of motion, we have used the linearization approximation, within which we can ignore the quadratic fluctuation terms like $\delta a^\dagger\delta a^\dagger$, $\delta a^\dagger\delta a$, e.t.c.
We now write the fluctuations in Eqs. B1(a-d) in terms of their Fourier components \citep{8},\\

\begin{subequations}
\begin{equation}
\begin{aligned}
\delta{a}(t)=A_1^-e^{-i\Omega t}+A_1^+e^{i\Omega t},
\end{aligned}
\end{equation}
\begin{equation}
\begin{aligned}
\delta{b}(t)=B_1^-e^{-i\Omega t}+B_1^+e^{i\Omega t},
\end{aligned}
\end{equation}
\begin{equation}
\begin{aligned}
\delta{x_1}(t)=q_1 e^{-i\Omega t}+q_1^*e^{i\Omega t}\hspace{0.2cm},
\end{aligned}
\end{equation}
\begin{equation}
\begin{aligned}
\delta{x_2}(t)=q_2 e^{-i\Omega t}+q_2^*e^{i\Omega t}\hspace{0.2cm}.
\end{aligned}
\end{equation}
\end{subequations}

This yields the following algebraic equations,

\begin{subequations}
\begin{equation}
\begin{aligned}
D_1A_1^-&=igB_1^--iG_1q_{1}\overline{a}-\sqrt{\kappa}\epsilon_p,
\end{aligned}
\end{equation}
\begin{equation}
\begin{aligned}
D_2A_1^+&=igB_1^+-iG_1\overline{a}q_1^*,
\end{aligned}
\end{equation}
\begin{equation}
\begin{aligned}
D_3B_1^-&=igA_1^--iG_2\overline{b}(q_2-q_1),
\end{aligned}
\end{equation}
\begin{equation}
\begin{aligned}
D_4B_1^+&=igA_1^+-iG_2\overline{b}(q_2^*-q_1^*),
\end{aligned}
\end{equation}
\begin{equation}
\begin{aligned}
\frac{q_1}{\chi_1(\Omega)}&=\hbar G_1(\overline{a}^*A_1^-+\overline{a}(A_1^+)^{\dagger})-\hbar G_2(\overline{b}^*B_1^-+\overline{b}(B_1^+)^{\dagger}),
\end{aligned}
\end{equation}
\begin{equation}
\begin{aligned}
\frac{q_2}{\chi_2(\Omega)}&=\hbar G_2(\overline{b}^*B_1^-+\overline{b}(B_1^+)^{\dagger}),
\end{aligned}
\end{equation}
\end{subequations}
where $D_1=\Theta_1+i \Omega$, $D_2=\Theta_1-i \Omega$, $D_3=\Theta_2+i \Omega$, $D_4=\Theta_2-i \Omega$.
Solving Eqs. B3(a-f) we get,\\
\begin{subequations}
\begin{equation}
\begin{aligned}
A_1^-&=\frac{gG_2(q_{2}-q_{1})\overline{b}-iD_{3}G_1q_{1}\overline{a}-D_{3}\sqrt{\kappa}}{D_1D_3+g^2},
\end{aligned}
\end{equation}
\begin{equation}
\begin{aligned}
B_1^-&=\frac{gG_1q_{1}\overline{a}-iD_1G_2(q_{2}-q_{1})\overline{b}-ig\sqrt{\kappa}\epsilon_c}{D_1D_3+g^2}.
\end{aligned}
\end{equation}
\end{subequations}

Similar equations for $A_1^+$ and $B_1^+$ can also be derived. To further simplify the calculations we assume that $D_{1}=D_{3}$ and $D_{2}=D_{4}$. This leads to the constraint, 
$\Theta_{1}=\Theta_{2}$. Since we are using only a single port output system, the standard input-output relation is,

\begin{subequations}
\begin{equation}
\begin{aligned}
a_{out}=a_{in}-\sqrt{\kappa}a(t)\hspace{3.1cm},
\end{aligned}
\end{equation}
\begin{equation}
\begin{aligned}
b_{out}=-\sqrt{\kappa}b(t)\hspace{3.7cm}.
\end{aligned}
\end{equation}
\end{subequations}

Using this we arrive at the output fields in the forward and backward directions,

\begin{subequations}
\begin{equation}
\begin{aligned}
a_{out}=C_{cb}e^{-i\omega_c t}+C_{pb}e^{-i\omega_p t}-\sqrt{\kappa}A_1^+e^{-i(2\omega_c-\omega_p)t},
\end{aligned}
\end{equation}
\begin{equation}
\begin{aligned}
b_{out}=C_{cf}e^{-i\omega_c t}+C_{pf}e^{-i\omega_p t}-\sqrt{\kappa}B_1^+e^{-i(2\omega_c-\omega_p)t}.
\end{aligned}
\end{equation}
\end{subequations}

Here $C_{cb}=\epsilon_c-\sqrt{\kappa}\overline{a}$, $C_{pb}=\epsilon_p-\sqrt{\kappa}A_1^-$, $C_{cf}=-\sqrt{\kappa}\overline{b}$, $C_{pf}=-\sqrt{\kappa}B_1^-$, are the complex coefficients. Using $C_{pb}$ and $C_{pf}$ we calculate the normalized backward reflection $T_b$ and forward transmission $T_f$ coefficients.\\

\begin{subequations}
\begin{equation}
\begin{aligned}
  T_{b}= |1- \kappa  [\frac{\frac{i g^2 G_{2}^2 \left(1-\frac{A}{B}\right)
   (1+C_{11}+C_{22})}{G_{1}}|\overline{b}|^2-i G_{1} D_{1}^2|\overline{a}|^2 -D_{1} g G_{2} \left( \left(1-\frac{A}{B}\right)\overline{a}^*\overline{b}+
   (1+C_{11}+C_{22})\overline{a}\overline{b}^*\right)}{\left(D_{1}^2+g^2\right)^2
   (C_{1}+C_{2}+C_{3})}\\+\frac{ig^2 G_{2} |\overline{b}|^2 }{B
   \left(D_{1}^2+g^2\right)^2}-\frac{D_{1}}{(D_{1}^2+g^2)}]|^2,
   \end{aligned}
\end{equation}
\begin{equation}
 \begin{aligned}
  T_{f}=|-\kappa  (\frac{g D_{1} \left(\frac{G_{2}^2\left(1-\frac{A}{B}\right)(1+C_{11}+C_{22})}  
   {G_{1}}|\overline{b}|^2+G_{1}|\overline{a}|^2 \right)-i G_{2} \left(g^2
   (1+C_{11}+C_{22})\overline{a}\overline{b}^*-D_{1}^2 \left(1-\frac{A}{B}\right)\overline{a}^*\overline{b}\right)}{\left(D_{1}^2+g^2\right)^2
   (C_{1}+C_{2}+C_{3})}\\+ \frac{g D_{1}G_{2}|\overline{b}|^2}{B \left(D_{1}^2+g^2\right)^2}-\frac{i
   g}{D_{1}^2+g^2})|^2,
 \end{aligned}
\end{equation}
\end{subequations}\\

where,
\begin{subequations}
\begin{equation}
 \begin{aligned}
 A=-\frac{g G_{1}\overline{a}\overline{b}^* +i D_{1} G_{2}|\overline{b}|^2 }{D_{1}^2+g^2}-\frac{g G_{1}\overline{a}^*\overline{b} -i  G_{2} D_{2}^*|\overline{b}|^2
  }{\left(D_{2}^*\right)^2+g^2},
\end{aligned}
\end{equation}
\begin{equation}
 \begin{aligned}
 B=\frac{iG_{2} D_{2}^* |\overline{b}|^2 }{\left(D_{2}^*\right)^2+g^2}-\frac{i D_{1}
   G_{2}|\overline{b}|^2 }{D_{1}^2+g^2}-\frac{1}{\hbar G_{2} {\chi_{2}} },
\end{aligned}
\end{equation}
\begin{equation}
 \begin{aligned}
 C_{1}=\frac{-g G_{2} \left(\left(1-\frac{A}{B}\right)\overline{a}\overline{b}^*+\overline{a}^*\overline{b}\right)+\frac{i G_{2}^2
   \left(1-\frac{A}{B}\right)|\overline{b}|^2 D_{2}^*}{G_{1}}+iG_{1}|\overline{a}|^2 D_{2}^*}{\left(D_{2}^*\right)^2+g^2},
\end{aligned}
\end{equation}
\begin{equation}
 \begin{aligned}
 C_{2}=\frac{-g G_{2} \left(\left(1-\frac{A}{B}\right)\overline{a}^*\overline{b} +\overline{a}\overline{b}^*\right)-\frac{iG_{2}^2\left(1-\frac{A}{B}\right)|\overline{b}|^2 D_{1} 
   }{G_{1}}-i G_{1}|\overline{a}|^2 D_{1} }{D_{1}^2+g^2},
\end{aligned}
\end{equation}
\begin{equation}
 \begin{aligned}
 C_{3}=-\frac{1}{ \hbar G_{1} \chi_{1}},
 \end{aligned}
\end{equation}
\begin{equation}
 \begin{aligned}
 C_{11}=\frac{g G_{1}\overline{a}^*\overline{b} +iG_{2} D_{1} |\overline{b}|^2}{B \left(D_{1}^2+g^2\right)},
 \end{aligned}
\end{equation}
\begin{equation}
 \begin{aligned}
 C_{22}=\frac{g G_{1}\overline{a}\overline{b}^* -i  G_{2} D_{2}^*|\overline{b}|^2}{B \left(\left(D_{2}^*\right)^2+g^2\right)}.
 \end{aligned}
\end{equation}
 \end{subequations}

\end{document}